\documentclass[twocolumn,aps,prl,floatfix,amssymb,superscriptaddress]{revtex4}

\usepackage{amsmath}
\usepackage{amssymb}
\usepackage{braket}
\usepackage{graphicx}
\usepackage{verbatim}
\usepackage{color}
\usepackage{amsthm}
 
\usepackage[unicode=true,pdfusetitle,
 bookmarks=true,bookmarksnumbered=false,bookmarksopen=false,
 breaklinks=false,pdfborder={0 0 0},backref=false,colorlinks=true]
 {hyperref}
 
\begin{document}

\title{Thermal Transport Signatures of Broken-Symmetry Phases in Graphene}

\author{Falko Pientka}
\author{Jonah Waissman}
\author{Philip Kim}
\author{Bertrand I. Halperin}
\affiliation{Department of Physics, Harvard University, Cambridge, Massachusetts 02138, USA}

\begin{abstract}
In the half-filled zero-energy Landau level of bilayer graphene, competing phases with spontaneously broken symmetries and an intriguing quantum critical behavior have been predicted. Here we investigate signatures of these broken-symmetry phases in thermal transport measurements. To this end we calculate the spectrum of spin and valley waves in the $\nu=0$ quantum Hall state of bilayer graphene. The presence of Goldstone modes enables heat transport even at low temperatures, which can serve as compelling evidence for spontaneous symmetry breaking. By varying external electric and magnetic fields it is possible to determine the nature of the symmetry breaking and temperature-dependent measurements may yield additional information about gapped modes.
\end{abstract}

\maketitle
{\em Introduction.}---The zero-energy Landau level of bilayer graphene has emerged as an intriguing experimental platform to study interaction-induced spontaneous symmetry breaking and competing orders. Following early theoretical work on quantum Hall ferromagnetism in graphene \cite{McCann2006,Yang2006,Aleiner2007,Barlas2008,Abanin2009,Lemonik2010,Lemonik2012,Nandkishore2012} several experiments have gathered evidence for a rich phase diagram of integer and fraction quantum Hall states in monolayer and bilayer graphene \cite{Feldman2009,Weitz2010,Martin2010,Velasco2012,Young2012,Maher2013,Young2014,Lee2014a,Maher2014,Kou2014,Hunt2016}.
Of particular interest are the $\nu=0$ states where short-range interactions on the lattice scale may lead to the formation of insulating phases and spontaneous symmetry breaking of either spin or isospin (valley) degrees of freedom \cite{Kharitonov2012,Kharitonov2012a}. Indeed, experimental observations based on electrical transport as well as capacitive measurements \cite{Maher2013,Young2014,Hunt2016} seem consistent with the phase diagram proposed in Refs.~\cite{Kharitonov2012,Kharitonov2012a}, however direct evidence for spontaneous symmetry breaking remains elusive.

As an alternative approach to charge based measurements, we propose to study broken-symmetry phases in the $\nu=0$ quantum Hall state of bilayer graphene by thermal transport. At low temperatures certain phases with spontaneous symmetry breaking can support heat flow via gapless spin or isospin waves, whereas heat conduction is blocked in phases without spontaneous order and a gapped spectrum of collective excitations. Hence thermal transport can yield direct evidence for spontaneous symmetry breaking and provide detailed information about collective modes. The technique is complementary to previous experiments as it is sensitive to both spin and isospin degrees of freedom and addresses insulating samples. In a broader context, thermal transport in graphene has recently been applied to study the breakdown of Fermi liquid behavior \cite{Crossno2016} as well as exotic electron-phonon scattering mechanisms \cite{Betz2013,Fong2013}.

\begin{figure}
\begin{minipage}{.175\textwidth}
\includegraphics[width=0.97\textwidth]{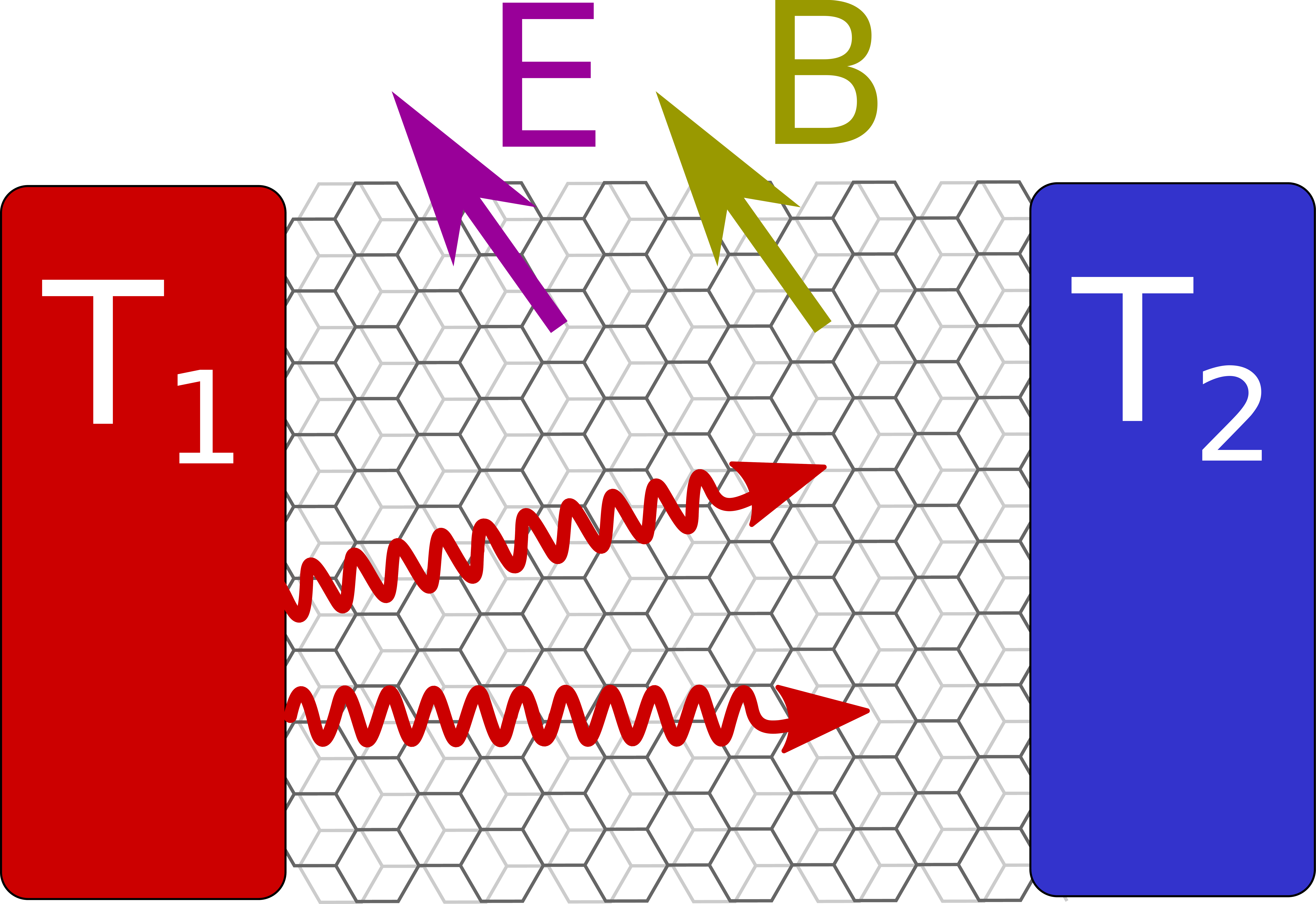}
\\[1.5em]
\includegraphics[width=0.97\textwidth]{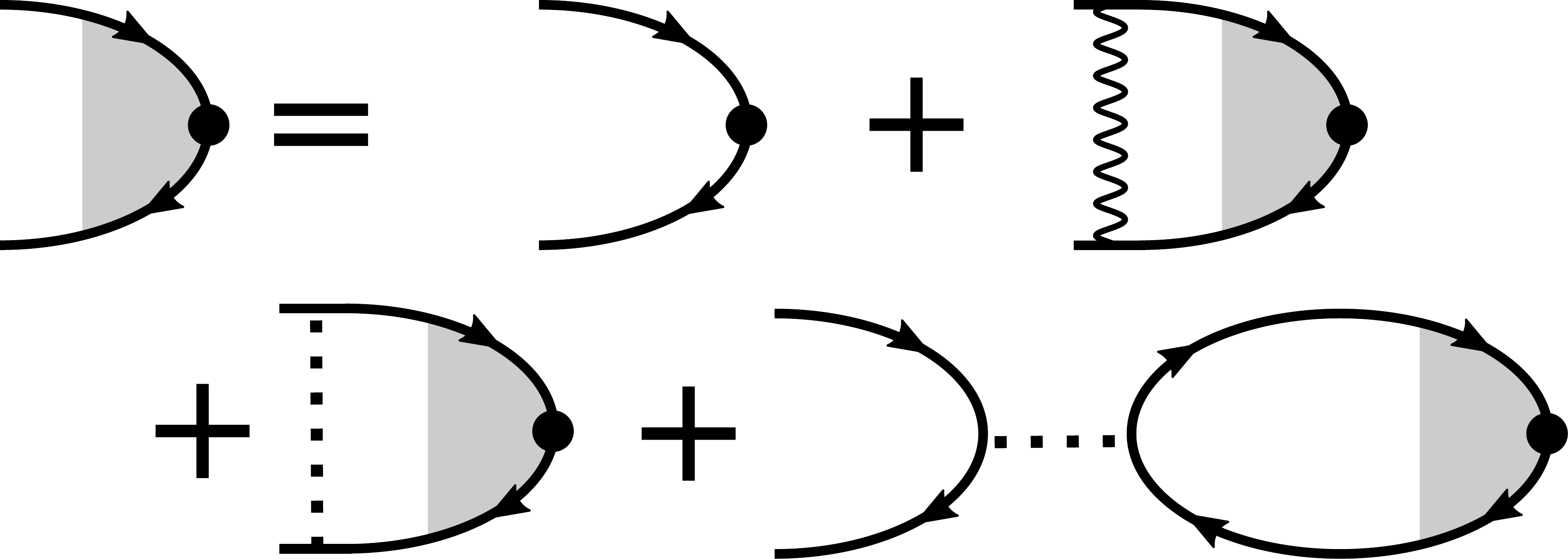}
\llap{\parbox[c]{1.0cm}{\vspace{-3.1cm}\hspace{-1.7cm}\footnotesize{(b)}}}
\llap{\parbox[c]{1.0cm}{\vspace{-7.6cm}\hspace{-1.9cm}\footnotesize{(a)}}}
\end{minipage}
\begin{minipage}{.3\textwidth}
\includegraphics[width=0.97\textwidth]{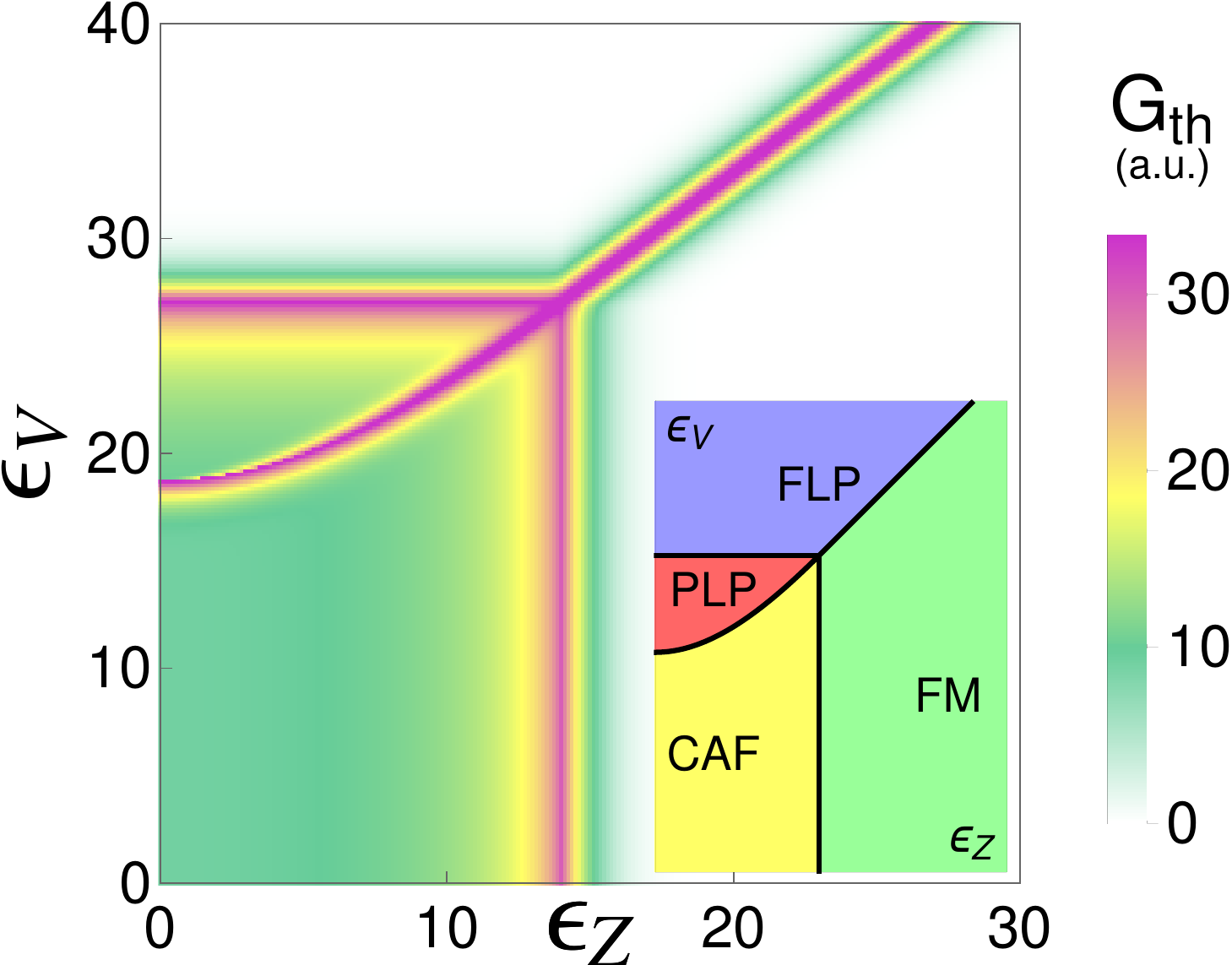}
\llap{\parbox[c]{1.0cm}{\vspace{-7.7cm}\hspace{-9.5cm}\footnotesize{(c)}}}
\end{minipage}
\caption{(a) Experimental setup: bilayer graphene in a perpendicular magnetic and electric field. Collective excitations transfer heat between two metallic leads at different temperatures. (b) Diagrammatic representation of the vertex function $\Gamma_{A,ab}$ comprising the Fock contribution of the long-range Coulomb interaction (wavy line) and Hartree and Fock contributions of the anisotropic short-range interaction (dotted line).
(c) Thermal conductance from particle-hole excitations at $T=1\,$K as a function of Zeeman energy and displacement field  \cite{foot1}. Inset: phase diagram of the $\nu=0$ state in bilayer graphene. The axes are the same as in the main plot.
}
\label{fig:setup}
\end{figure}

A possible experimental setup is shown in Fig.~\ref{fig:setup}(a). A graphene bilayer is contacted by leads at different temperatures which can be measured separately as in recent experiments \cite{Fong2012,Fong2013,Crossno2015}. A strong perpendicular magnetic field drives the system into a quantum Hall regime and induces a Zeeman energy $\epsilon_Z$. While interactions in graphene quantum Hall states typically depend on the perpendicular magnetic field, the Zeeman energy depends on the total magnetic field and its magnitude can be tuned independently by applying an additional in-plane field. A perpendicular electric field induces a displacement field $\epsilon_V$ between the two layers. In the zero-energy Landau level, the layers are coupled to the isospin degrees of freedom and thus $\epsilon_{Z/V}$ control the spin (isospin) polarization. Before introducing the details of our model we briefly outline the central result of our work. 
Figure~\ref{fig:setup}(c) shows the contribution to the low-temperature thermal conductance $G_{\rm th}$ from collective modes as a function of the Zeeman energy and displacement field. Comparison with the phase diagram in the inset reveals that thermal transport exhibits clear signatures of spontaneous symmetry breaking. Heat can flow via Goldstone modes inside the canted antiferromagnet (CAF) or partially-layer polarized (PLP) phase due to spontaneously broken spin or isospin symmetries. In contrast, the thermal conductance from collective modes vanishes in the fully layer polarized (FLP) and ferromagnetic (FM) phases, where the excitation spectrum is gapped \cite{foot1}. Moreover $G_{\rm th}$ is dramatically enhanced near phase transitions, due to the large low-energy density of states of gapless quadratic modes present at the critical points.

{\em Model.}---We follow Refs.~\cite{Kharitonov2012,Kharitonov2012a} to describe symmetry breaking in the lowest Landau level of bilayer graphene in the Hartree--Fock approximation. We assume Landau level spacing to be large and project onto the lowest Landau level. In this approximation, we cannot describe the system in terms of bare Coulomb interactions, which can be strongly renormalized by Landau level mixing even at strong fields \cite{Kharitonov2012}. Instead we assume renormalized parameters which are determined from experiment.

The zero-energy Landau level of bilayer graphene is eightfold degenerate \cite{McCann2006}. The states are annihilated by operators $c_{a,k}$ with index $a=\{n_a,\alpha\}$ combining orbital index $n_a=0,1$ and SU(4) index $\alpha$ that accounts for spin and isospin.
We shall use the basis  $\Psi_{a}({\bf r})=\sum_k\braket{{\bf r}|\psi_{a,k}}c_{a,k}$, where $\ket{\psi_{a,k}}=\chi_{\alpha} \ket{n_a,k}$ with $\chi_{\alpha}$ an SU(4) spinor and $\ket{n,k}$ the Landau level wavefunctions.

The Hamiltonian is $H=\int d^2r \Psi_{a}^\dag({\bf r}){\cal H}_{s,\alpha}\Psi_{a}({\bf r})+H_i$, where ${\cal H}_s=-\epsilon_Z\sigma_z-\epsilon_V\tau_z$ is the single-particle term, $\sigma_j$ ($\tau_j$) are Pauli matrices in spin (isospin) space, and $H_i=(1/2)\sum_{abcd}\int d^2rd^2r' \Psi^\dag_a({\bf r})\Psi^\dag_b({\bf r}')\Psi_c({\bf r}')\Psi_d({\bf r}) V_{\alpha \beta \gamma \delta}({\bf r}-{\bf r}')$ with $ V_{\alpha \beta \gamma \delta}({\bf r})=(e^2/\epsilon r)\delta_{\alpha \delta}\delta_{\beta \gamma }+ \delta({\bf r})W_{\alpha \beta \gamma \delta}$ describes Coulomb interactions. Introducing the parameters $u_x=u_y=u_\perp$ and $u_z$ the short-range contribution can be expressed as $W_{\alpha \beta \gamma \delta}=\pi l_B^2\sum_{j=x,y,z} u_j(\tau_j)_{\alpha\delta}(\tau_j)_{\beta\gamma}$, with $l_B$ the magnetic length \cite{Kharitonov2012,Kharitonov2012a}. Our model neglects terms of higher order in the interaction which break the U(1) isospin symmetry down to a discrete $C_3$ symmetry due to the lattice structure of graphene. 

\begin{figure}
\includegraphics[width=0.235\textwidth]{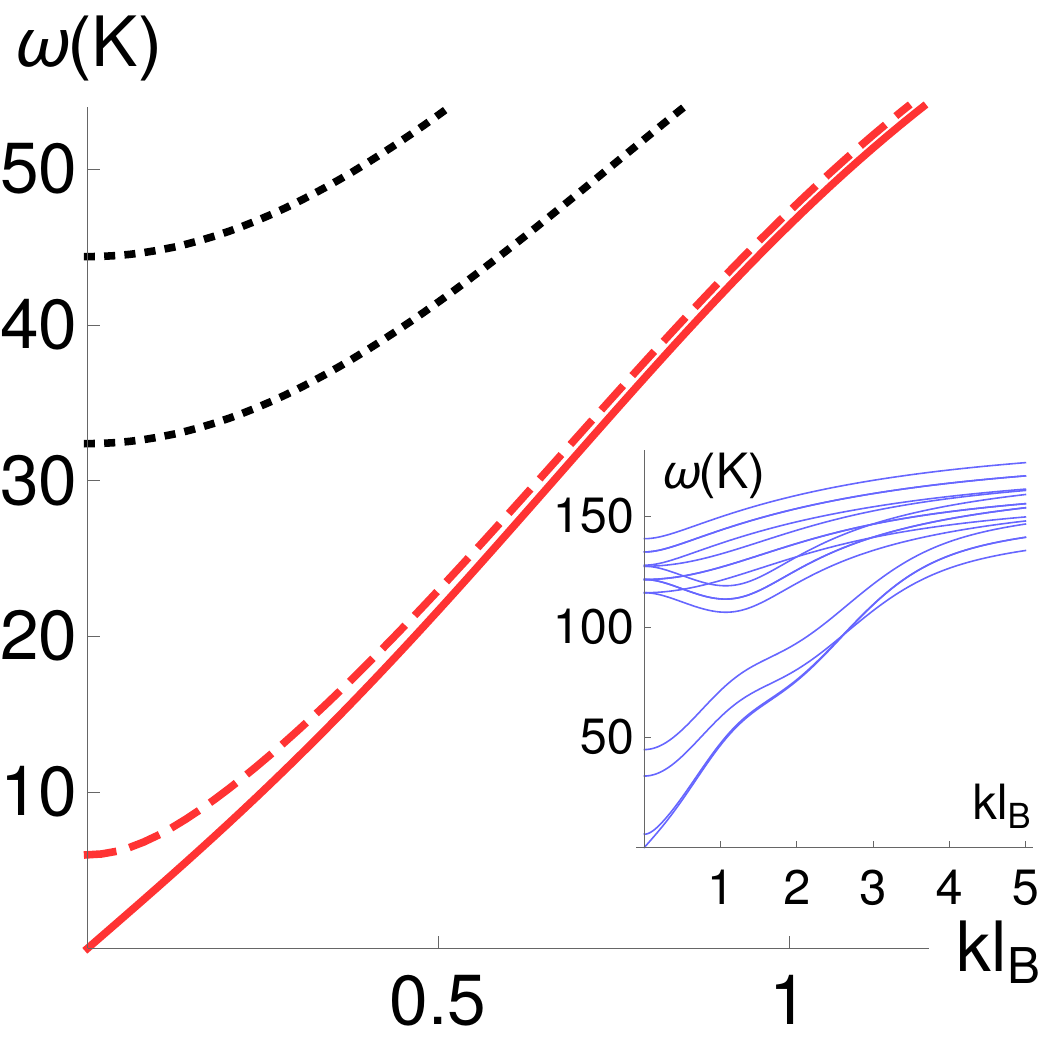}
\includegraphics[width=0.235\textwidth]{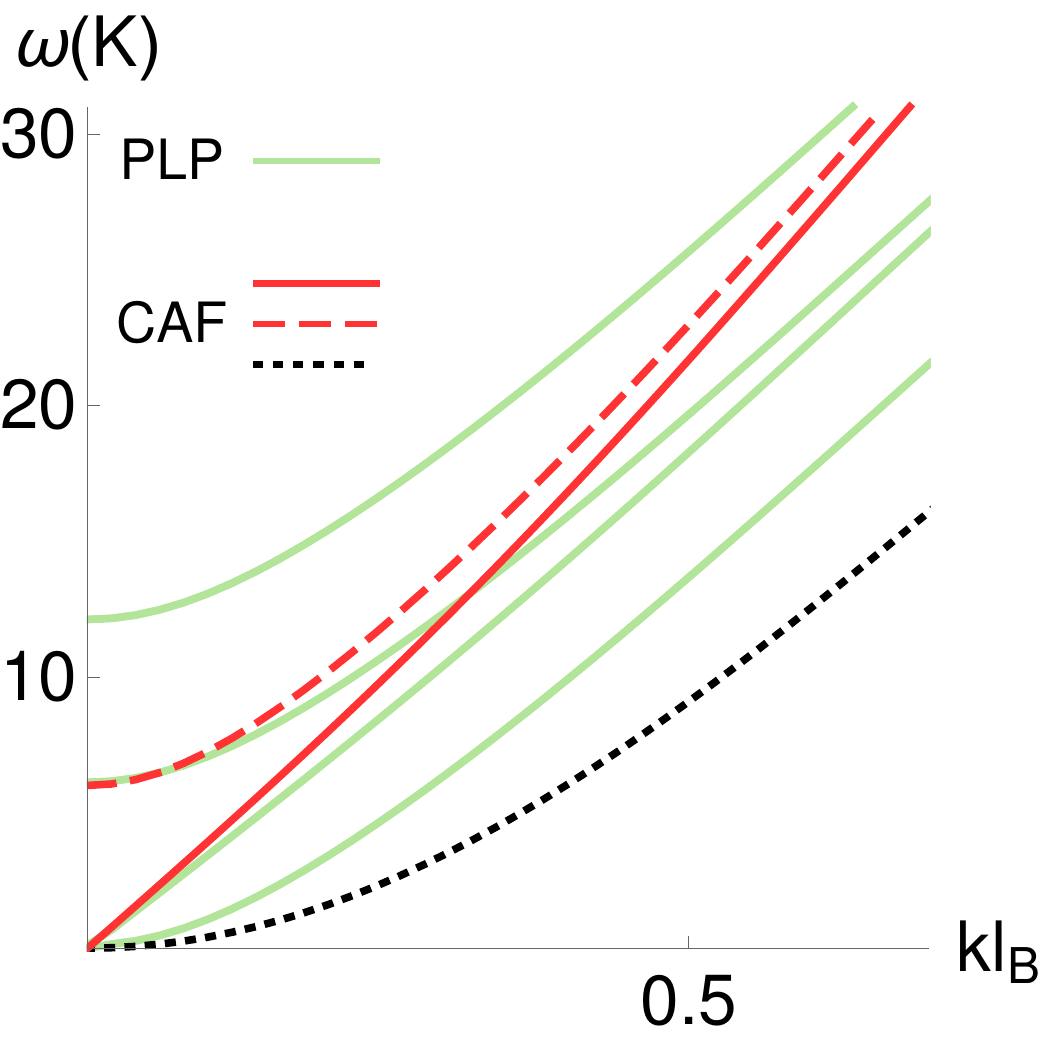}
\hspace{-1em}
\llap{\parbox[c]{1cm}{\vspace{-7.8cm}\hspace{-16.2cm}\footnotesize{(a)}}}
\llap{\parbox[c]{1cm}{\vspace{-7.8cm}\hspace{-7.9cm}\footnotesize{(b)}}}
\caption{(a) Excitation spectrum in the CAF phase for $\epsilon_V=3\,$K. The two low-energy modes (red solid and dashed) are spin waves at low wavevectors, while the higher-energy modes (black dotted) correspond to fluctuations in the isospin sector. Inset: full spectrum with energies up to $\sim U_c$. (b) Excitation spectrum at $\epsilon_V= 19.2\pm 0.1\,$K on the two sides of the CAF--PLP transition. The spectrum on the CAF side is shown in the same color coding as in (a) and the spectrum on the PLP side is shown as green solid lines.}
\label{fig:spectrum}
\end{figure}

\begin{figure}[t]
\includegraphics[width=0.45\textwidth]{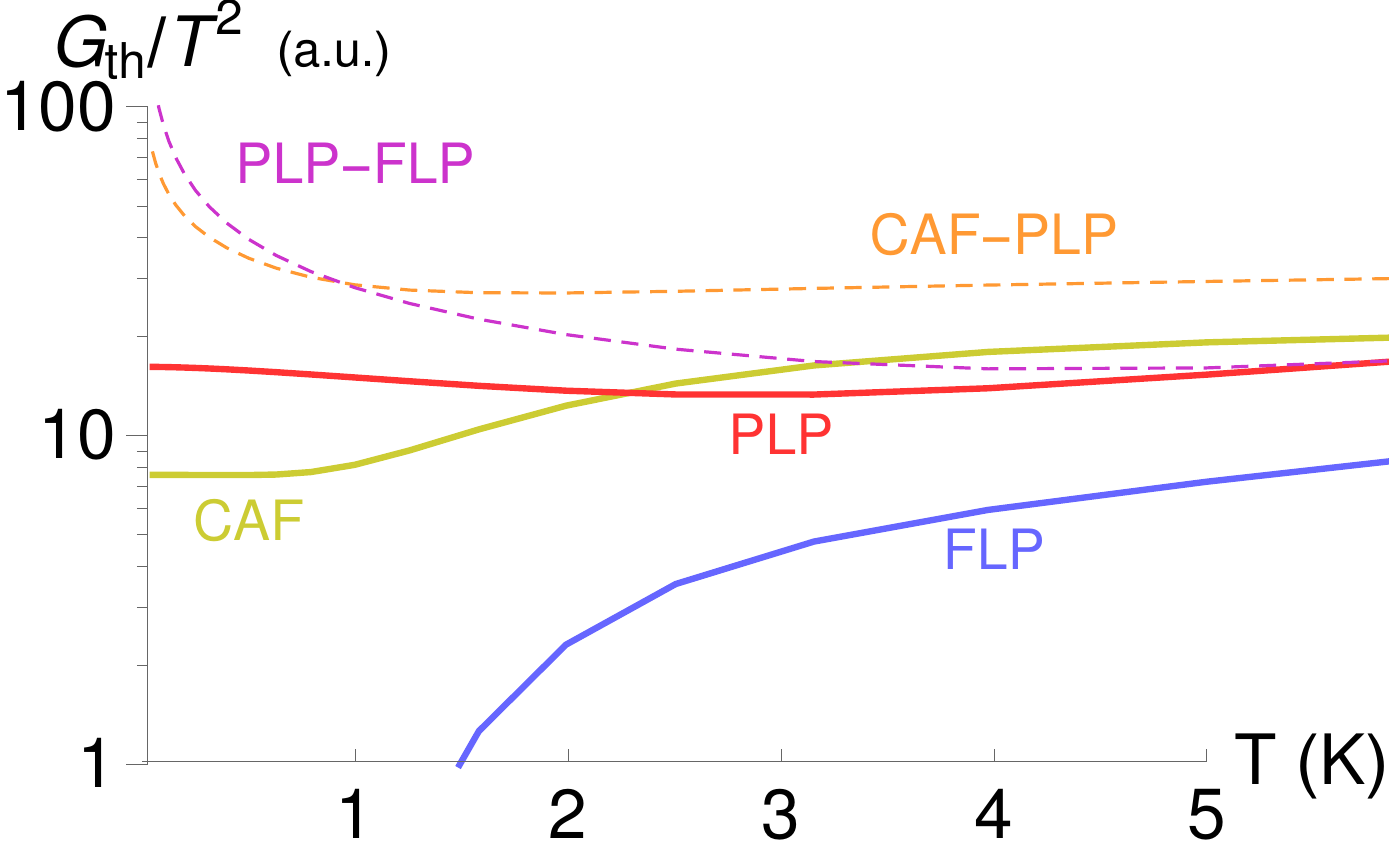}
\llap{\parbox[c]{1.0cm}{\vspace{-9.6cm}\hspace{-15.6cm}\footnotesize{(a)}}}\\
\includegraphics[width=0.45\textwidth]{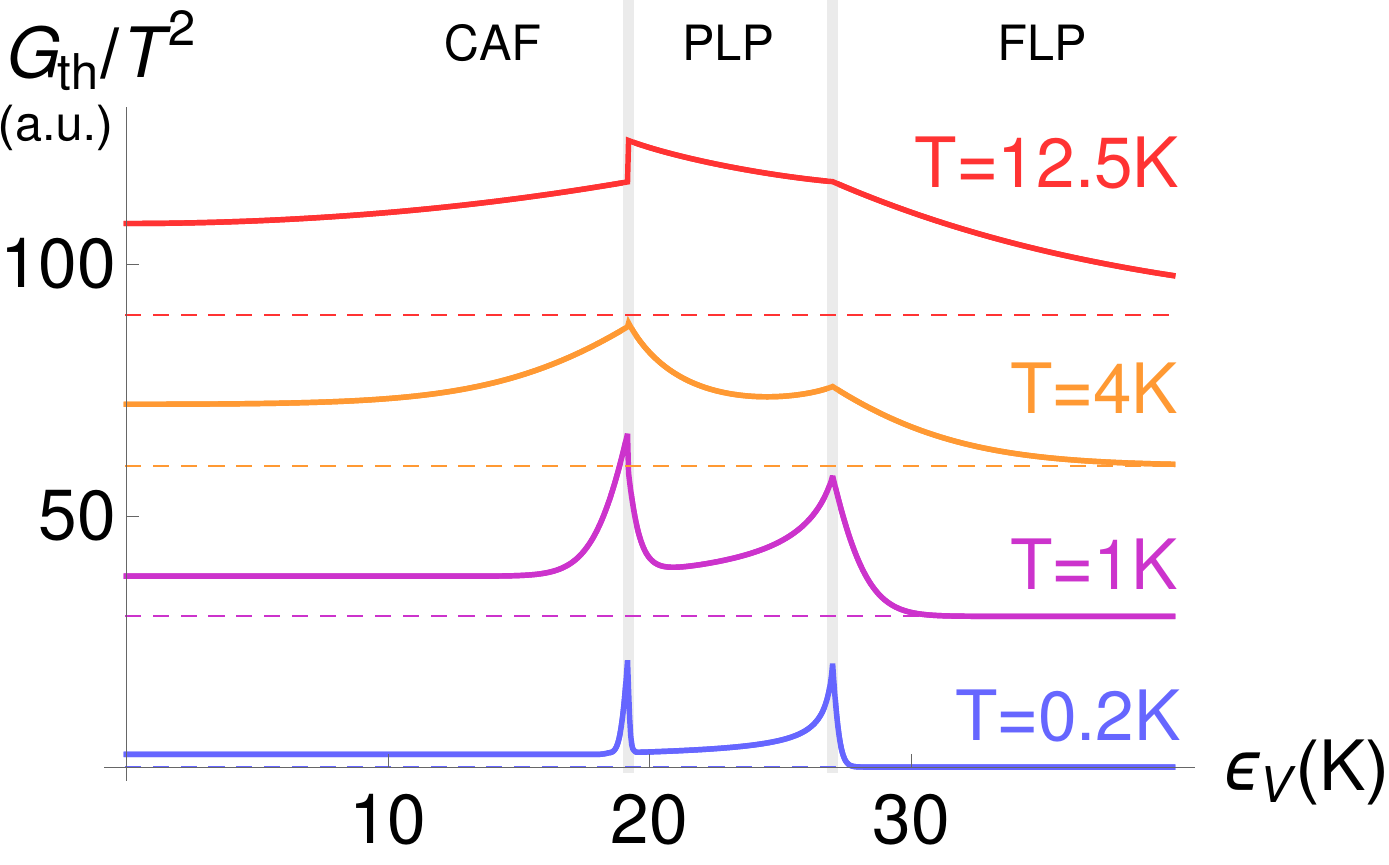}
\llap{\parbox[c]{1.0cm}{\vspace{-10cm}\hspace{-15.6cm}\footnotesize{(b)}}}
\caption{(a) Thermal conductance normalized by $T^2$ inside the three phases CAF, PLP, and FLP at $\epsilon_V=15\,$K, $25\,$K, $31\,$K (solid lines) as well as near the phase transitions CAF--PLP and PLP--FLP at $\epsilon_V=19.21\,$K, $27\,$K (dashed lines). (b) Thermal conductance {\em vs} $\epsilon_V$ for various temperatures. Curves have been offset for clarity (the dashed lines indicate zero conductance) and the vertical gray lines mark the phase transitions.}
\label{fig:condV}
\end{figure}

Guided by experiment \cite{supp}, we shall use the following parameters appropriate for a perpendicular magnetic field of $B_\perp=5{\rm T}$: $\epsilon_Z=3\,$K (Figs.~\ref{fig:spectrum} and \ref{fig:condV}), $u_\perp=-7 \,$K, $u_z=20\,$K, and $U_c=100\,$K. Assuming a sample width of $W=5\,\mu$m, we estimate a thermal conductance of the Goldstone mode in the CAF phase of $G_{\rm th}=17\,$pW/K at $T=1\,$K  \cite{supp}, which is accessible in experiment \cite{Fong2013}.

{\em Excitations within the $\nu=0$ Landau level.}---We calculate the spectrum of intra-Landau-level particle-hole excitations following Ref.~\cite{Kallin1984} from the response to an operator $\theta_A$ in SU(4) spin space
\begin{align}
\chi_A({\bf k},&\omega)=\sum_{ab}\sum_{q_1,q_2}\braket{q_1a|e^{i{\bf kr}}\theta_A^\dag|q_2b}\nonumber\\
\times&\int \frac{d\omega'}{2\pi}G_a(\omega+\omega')G_b(\omega')\Gamma_{A,ab}(q_1,q_2;\mathbf{k},\omega),
\end{align}
where the vertex part $\Gamma_{A}$ is given by the diagrams shown in Fig.~\ref{fig:setup}(b). The excitation spectrum can be found from the poles of the response function $\chi_A$. A tedious but straightforward calculation \cite{supp} yields the equation
\begin{align}
 \sum_{cd}[&\delta_{ac}\delta_{bd} 
 [(f_c-f_d)\omega-E_{{\rm xc},cd}]-\tilde{V}_{abcd}({\bf k})]
B_{cd}=0\label{eval_problem},
 \end{align}
where $E_{{\rm xc}}$ is the exchange energy, which accounts for single-particle and self energy contributions, and 
$\tilde{V}_{abcd}$ denotes interaction matrix elements comprising Hartree and Fock contributions \cite{supp}. This equation has to be satisfied for all pairs $(a,b)$ and $(c,d)$ with one occupied and one empty state. At low energies the resulting $32\times 32$ eigenvalue problem can further reduced \cite{supp} to an $8\times 8$ problem in the limit of large Coulomb energy $U'_c=N_0U_c\gg u_\perp,u_z$ with $U_c=\sqrt{\pi/2}(e^2/\epsilon l_B)$ and $N_0=89/224$. Incidentally, the resulting eigenvalue equation has the same analytical form as for monolayer graphene when $\epsilon_V=0$ and $N_0=1/4$. Hence, our results can also be applied to monolayers (see also \cite{Wu2014}).

The $8\times 8$ matrix can be readily diagonalized and we obtain four low-energy modes along with their particle-hole symmetric partners.
We expand the spectrum at long wavelength, $kl_B\ll 1$, and find in the CAF phase, where $u_\perp<0$,
\begin{align}
 \omega_{1} \simeq &v k, \qquad 
 v=2l_B\sin\theta_s\sqrt{|u_\perp| U'_c},\label{spin_goldstone}\\
 \omega_{2} \simeq &2\epsilon_Z + k^2l_B^2U'_c(\epsilon_Z^2+4 u_\perp^2)/4 |u_\perp| \epsilon_Z ,\label{CAF2}\\
 \omega_{3,4} \simeq &\Delta_c\mp 2\epsilon_V+k^2l_B^2U'_c(\epsilon_Z+4|u_\perp| u_z)/2|u_\perp|\Delta_c\label{CAF3}
\end{align}
with $\Delta_c^2=4(u_z+|u_\perp|)(u_z-|u_\perp|+\epsilon_Z^2/2|u_\perp|)$. The spectrum is shown in Fig.~\ref{fig:spectrum}(a). We find a gapless mode in agreement with the spontaneously broken $U(1)$ spin symmetry in the CAF phase and the next higher mode with a gap of $2\epsilon_Z$, which corresponds to fluctuations of the spin component along the Zeeman field direction. The higher modes $\omega_{3/4}$ are associated with fluctuations in the isospin sector. The velocity of the Goldstone mode depends on the canting angle $\cos\theta_s=\epsilon_Z/2|u_\perp|$.

Increasing the Zeeman field polarizes the spins along the field direction until the system enters the FM phase when $\theta_s=0$ and the velocity of the Goldstone mode vanishes. The gapless modes thus becomes quadratic at the critical point and a gap opens as the system enters the FM phase \cite{supp} and spin fluctuations around the Zeeman field direction are further suppressed by an increasing field strength.

If, instead, we increased the displacement field $\epsilon_V$, the CAF phase transitions into the PLP phase when $\epsilon_V=\Delta_c/2$ \cite{Kharitonov2012a}. At this point the quadratic mode $\omega_3$ is gapless in addition to $\omega_1$. Interestingly the excitation spectrum is discontinuous at the critical point reflecting the first-order nature of this transition. The spectrum in the PLP phase for $kl_B\ll 1$ is given by \cite{supp}
\begin{align}
\omega_1\simeq &vk, \ v\simeq l_B\sin\theta_m\sqrt{2 U'_c(u_z+|u_\perp|)},\label{isospin_goldstone}\\
\omega_{2}\simeq&\Delta_p +k^2l_B^2 U'_c [(u_z+|u_\perp|)\sin^2\theta_m - 4 |u_\perp|]/\Delta_p ,\label{PLP2}
\end{align}
and $\omega_{3,4}=\omega_2\mp 2\epsilon_Z$. Here we have introduced the canting angle of the isospin $\cos\theta_m=\epsilon_V/(u_z+|u_\perp|)$ and $\Delta_p^2=8|u_\perp|(\epsilon_V^2 + u_\perp^2 - u_z^2)/(u_z+|u_\perp|)$. The PLP phase features a Goldstone mode $\omega_1$ due to the broken U(1) symmetry in the isospin sector, along with several gapped modes. If the CAF--PLP transition is approached from the PLP side, $\omega_3$ becomes gapless. The spectra on the CAF and PLP side of the transition are compared in Fig.~\ref{fig:spectrum}(b). The three lowest modes coincide in both phases at $k=0$, but have different dispersions.
Interestingly, another low-energy mode with a gap of $4\epsilon_Z= 12\,$K is present in the PLP phase. In contrast, the fourth mode in the CAF is shifted to much higher energies, $4\epsilon_V\simeq 77\,$K.

Further increasing $\epsilon_V$ leads to a stronger polarization of the isospins and the system enters the FLP phase once $\theta_m=0$. This transition is continuous and the gapless modes becomes quadratic at the critical point similar to the CAF--FM transition. On the FLP side this mode acquires a gap which grows with increasing displacement field as isospin fluctuations are suppressed \cite{supp}.

{\em Thermal conductance.}---Thermal transport at low temperatures can indicate the presence of neutral low-energy excitations. Importantly, the broken-symmetry phases CAF and PLP support thermal transport via Goldstone modes and can thus be readily distinguished from phases with a gapped spectrum and exponentially suppressed thermal conductance at low temperatures. In order to relate the thermal conductance to the presence of particle-hole excitations we must rule out other origins of heat transport, most importantly, electrons and phonons.

At low temperatures we can safely ignore electronic heat transport in the insulating phases CAF, PLP, and FLP \cite{foot3}. While the FM phase is also insulating in the bulk, it has conducting edge states \cite{Maher2013,Young2014}, which could contribute to heat transport \cite{foot4}. Such an electronic heat current can in principle be suppressed by additional side leads acting as heat sinks for the edge states.

Heat transport via phonons in nonsuspended samples can presumably be neglected in the setup of Fig.~\ref{fig:setup}(a). Phonons will thermalize with the substrate before reaching the drain and thus energy transfered to phonons is lost. We expect such losses to be small at low temperatures due to the high power-law decay of the electron-phonon energy relaxation rate $\propto T^4$ in graphene \cite{Betz2013,Crossno2015}.  

In the following we focus on the heat current carried by particle-hole excitations and ignore other contributions. We model the leads as black bodies assuming the interface between leads and sample to be sufficiently rough. We furthermore assume ballistic transport through the sample ignoring defects and scattering between SU(4) spin waves. Ballistic transport of magnons on a millimeter scale has been observed in the three-dimensional antiferromagnet Nd$_2$CuO$_4$ \cite{Li2005}. In contrast, earlier works on quasi-two-dimensional antiferromagnets found typical mean free paths of $\sim100\,$nm attributed to scattering off defects \cite{Sologubenko2000,Sales2002,Hess2003}. However, it is reasonable to assume a considerably longer scattering length in high-quality graphene samples, where such defects are not present. The magnon-magnon scattering rate in a two dimensional antiferromagnet for modes with $vk\sim T$ is $\sim vk (T/2\pi\rho)^2$ \cite{Tyc1990}, where $\rho\sim U_c'$ is the spin stiffness. In the CAF phase at $T=10\,$K this yields a magnon-magnon scattering length of several microns.

In two-dimensions the heat current through a medium of width $W$ with an isotropic dispersion $\omega(k)$ is given by \cite{Ziman}
$ J =(W/2\pi^2)\int dk \, k \omega(k) v_k  n_B[\omega(k)]$, where $v_k=\partial_k\omega(k)$ and $n_B(\omega)$ is the Bose distribution. We assume the sample to be much longer than the thermal wavelength such that we can neglect evanescent modes and resonances. We then obtain the thermal conductance as $G_{\rm th}=(J_L-J_R)/\Delta T$ where $J_{L/R}$ are the heat currents from the left and right reservoir with temperature $T_{L/R}$ and $\Delta T=|T_L-T_R|\ll T_L$.

The thermal conductance of the different phases can be compared in a single sample by tuning $\epsilon_V$ via a gate voltage and $\epsilon_Z$ by an in-plane magnetic field. Figure~\ref{fig:setup}(c) shows the thermal conductance at $T=1\,$K along with the corresponding phase diagram in the inset. The phases with broken symmetries, CAF and PLP, exhibit thermal transport, whereas the conductance is exponentially suppressed inside the gapped phases, FLP and FM.

The phase transitions give rise to particularly strong signatures in thermal transport. At the first-order transition between the CAF and PLP phases a mode with a quadratic dispersion becomes gapless in addition to the linear Goldstone modes present in both phases [cf.\ Fig.~\ref{fig:spectrum}(b)]. Thus once the energy minimum of the quadratic mode becomes comparable to temperature, the available phase space for heat conduction increases dramatically. The continuous transition between PLP and FLP does not involve any additional low-energy modes, but the linear dispersion of the gapless mode in the PLP phase becomes quadratic at the critical point. The large density of states at the bottom of the quadratic band causes $G_{\rm th}$ to peak at the transition.

We emphasize that all phase transitions have prominent signatures in the thermal conductance. This constitutes an important result of our work.
While previous experiments based on charge transport have identified phase boundaries \cite{Weitz2010,Velasco2012,Maher2013} it remains unclear which transitions should exhibit signatures in such experiments \cite{Lee2015,Dhochak2015}. 
Thus thermal transport measurements could possibly lead to the discovery of previously unresolved phase transitions. Moreover, they could determine whether the layer-polarized state observed in experiment is a broken-symmetry PLP state or a gapped FLP state.

While low-temperature measurements are sufficient to determine the phase diagram, more detailed information about the excitation spectrum can be gained by varying temperature. The conductance of a linear dispersion, $\omega(k)=vk$, follows the Stefan--Boltzmann law in two dimensions $G_{\rm th}\sim WT^2/v$. In contrast, a gapless quadratic mode leads to a weaker power-law dependence $G_{\rm th}\propto T^{3/2}$ \cite{supp}. The conductance of a gapped spectrum is exponentially suppressed at low temperatures.

In the following we concentrate on the three electronically insulating phases CAF, PLP and FLP. Figure~\ref{fig:condV}(a) shows the temperature dependence of the conductance divided by $T^2$ within the three phases and at the transition points. The thermal conductance in FLP shows an activated behavior from which one can extract the gap in the excitation spectrum. The conductance in the PLP phase follows approximately a $T^2$ law indicating that only the linear Goldstone mode has significant occupation up to a temperature of $T=5\,$K. In the CAF phase $G_{\rm th}\propto T^2$ at low temperatures $T\lesssim 1\,$K indicating the presence of a Goldstone mode. 
At temperatures above $3\,$K the conductance follows a $T^2$ law with a larger proportionality constant, which implies that additional gapped low-energy excitations become occupied above $T=1\,$K.
At the critical points between CAF, PLP and FLP, the conductance increases slower than $T^2$ and thus $G_{\rm th}/T^2$ decreases. This is a signature of gapless quadratic modes with a characteristic $T^{3/2}$ behavior. Hence the conductance peaks at phase transitions become particularly pronounced at low temperatures.

Additional information can be obtained from traces of the conductance as a function of displacement field for fixed Zeeman energy shown in Fig.~\ref{fig:condV}(b). Interestingly, the signature of the CAF--PLP transition changes with temperature. A strong peak at low-temperatures turns into a pronounced discontinuity above $T=10\,$K. This is a direct signature of a first-order phase transition, where the excitation spectrum changes discontinuously between the two phases as shown in Fig.~\ref{fig:spectrum}(b). The most striking difference between the two spectra is an additional mode in the PLP phase with an energy minimum of $12 \,$K, which causes a sizable jump in the conductance. This discontinuity survives with increasing temperature until high-energy modes at around $\omega=70\,$K get populated.

The PLP--FLP transition, in contrast, is of second-order and thus the conductance is continuous. Instead, the derivative of the conductance with respect to $\epsilon_V$ jumps at the critical point, signaling a higher-order phase transition.
The FLP phase shows activated transport whose activation energy grows with distance from the critical point. This is a clear signature of the FLP phase where the gap increases with layer polarization.

The conductance as a function of displacement field can also serve to distinguish CAF and PLP phases and clarify the character of the gapless modes. In the CAF phase the lowest-energy excitations are spin waves, which leave the mean isospin polarization unchanged. Hence the Goldstone mode velocity does not depend on the displacement field which acts only on isospin degrees of freedom and the conductance remains constant in a broad range of fields. In the PLP phase the Goldstone mode is an isospin wave and hence its velocity decreases with $\epsilon_V$ in agreement with Eq.~(\ref{isospin_goldstone}). Consequently, the thermal conductance grows when increasing $\epsilon_V$ according to the Stefan--Boltzmann law, $G_{\rm th}\propto 1/v$. This distinctive behavior is clearly visible at $T=0.2\,$K and $1\,$K in Fig.~\ref{fig:condV}(b).

Alternatively, one can verify the magnetic nature of the Goldstone mode in the CAF phase by tuning the Zeeman field, which changes the spin-wave velocity but leaves the isospin-wave velocity in the PLP phase unchanged. Thus by tuning $\epsilon_V$ and $\epsilon_Z$ one can elucidate the nature of the broken symmetry in the ground state and the spin-isospin structure of the Goldstone modes.

{\em Conclusions.}---Low-temperature thermal transport in the $\nu=0$ state of bilayer graphene exhibits unique signatures of various ordered phases and phase transitions. Obtaining direct evidence of the CAF and PLP states would be particularly intriguing as the first-order transition between theses states closely resembles the anti\-ferro\-magnet--superconducting transition in some high $T_c$ superconductors \cite{Wu2014,Lee2014,Lee2015} and may exhibit deconfined quantum criticality \cite{Senthil2004}.
It would be interesting to extend the present work to study other insulating phases in two dimensions such as in moir\'e heterostructures \cite{Cao2016}. Another exciting future direction is to study the bulk thermal transport of broken-symmetry quantum Hall states at other filling factors which could possibly be achieved by introducing heat sinks for the conducting edge channels. 

During the final stages of this work we became aware of Ref.~\cite{deNova2017} which also calculates the excitation spectrum of bilayer graphene.

We acknowledge financial support by the STC Center for Integrated Quantum Materials, NSF Grant No.\ DMR-1231319.

\bibliography{caf}

\begin{thebibliography}{45}
\bibitem[{foo({\natexlab{c}})}]{foot4}
\bibinfo{note}{The ballistic edge states of the FM phase would contribute a
  thermal conductance of $ 4\,{\rm pWK}^{-2}\times T$.}
\end{thebibliography}

\clearpage

\onecolumngrid

\section{Thermal conductance in two dimensions}

Assuming black-body radiation, the thermal conductance can be obtained from \cite{Ziman}
\begin{align}
 G_{\rm th}=\frac{d}{dT}\frac{W}{\pi}\int d\omega n_B(\omega)\nu(\omega)\omega\partial_k\omega.
\end{align}
Using $\nu(\omega)=k/2\pi\partial_k\omega$ one obtains the expression given in the main text, $G_{\rm th}=\frac{d}{dT}(W/2\pi^2)\int dk  n_B[\omega(k)] \omega(k)k\partial_k\omega$. For a gapped linear spectrum, $\omega=\sqrt{\Delta^2+v^2k^2}$, the thermal conductance is given by 
\begin{align}
 G_{\rm th} &=\frac{3WT^2}{2\pi^2v }f(\Delta/T) ,
\end{align}
where we have defined $f(z)=\int^\infty_z dx x \sqrt{x^2-z^2}/(e^{x}-1)$ and we have set $k_B=1$. The function $f$ decays monotonically and can be approximated for large arguments by $f(z)\simeq e^{-z}\sqrt{z^3\pi/2}$. At temperatures $T=\Delta/4$ the conductance is reduced by an order of magnitude. For a gapless dispersion, $\Delta=0$, we recover the Stefan--Boltzmann relation in two dimensions
\begin{align}
 G_{\rm th}(\Delta=0) &=\frac{3\zeta(3) WT^2}{\pi^2v }\label{stefan-boltzmann}.
\end{align}
A quadratic dispersion, $\omega(k)=k^2/2m$, has a weaker temperature dependence, instead,
\begin{align}
 G_{\rm th} &=\frac{9\zeta(5/2)}{4\sqrt{2\pi^3}}W \sqrt{m}T^{3/2}.
\end{align}

\section{physical parameters}

We estimate the physical parameters of our model mainly from the experiment in Ref.~\cite{Maher2013} for a perpendicular magnetic field of $B_\perp=5\,{\rm T}$. We emphasize, however, that physical parameters may vary between different experiments in particular as screening by nearby gate electrodes can influence the effective interactions.
The parameter $u_\perp$ can be obtained from the critical Zeeman field at the CAF-FM transition, when magnetic field is rotated. For a perpendicular field $B_\perp=5\,$T the critical total field has been determined as $B_{\rm tot}^c=21 \,$T \cite{Maher2013}. With a Zeeman energy $\epsilon_Z/B_{\rm tot}=0.7\,{\rm KT}^{-1}$ this yields an anisotropy energy $u_\perp(B_\perp=5\,{\rm T})=\epsilon_Z/2=7.3 \,$K. In a total field of $B_{\rm tot}=5\,$T we have a Zeeman energy of $\epsilon_Z=3.5\,$K which yields a canting angle $\theta\sim 0.4\pi$. The anisotropy energy $u_z$ can be estimated from the phase transition to the layer polarized phase $u_\perp+u_z=\epsilon_V^2/(u_z-u_\perp)+\epsilon_Z^2/2u_\perp$, where $\epsilon_V=e{\cal E} a_z/2$ with ${\cal E}$ the electric field and $a_z=3.5\,$\AA{} the interlayer distance. From \cite{Maher2013} we obtain  $u_z/B_\perp\simeq 14-20\,{\rm K T}^{-1}$. For the purpose of presentation it is useful to choose a somewhat lower value here, $u_z=20\,$K, although our results remain qualitatively valid for higher values of $u_z$.

The bare Coulomb energy $e^2/l_B= 650 \,{\rm K}\sqrt{B_\perp}$, where $B_\perp$ is measured in Tesla, can be screened by an electronic polarization in graphene or by a substrate or gate electrodes \cite{Kharitonov2012}. We therefore assume a renormalized Coulomb energy $U_c=e^2/\epsilon l_B$ with an effective dielectric constant $\epsilon$ that takes all these effects into account. Its value can be estimated from the single-particle gap in the system and may vary from sample to sample. For the $\nu=0$ state in suspended bilayer graphene Ref.~\cite{Martin2010} reports a gap of $100\,$K at $B_\perp=5\,$T and we therefore set $U_c=e^2/\epsilon l_B=100\,$K, which corresponds to an effective dielectric constant of $\epsilon\simeq 15$.

With these parameters we obtain a spin wave velocity in the CAF phase
\begin{align}
 v=l_B\sin\theta\sqrt{|u_\perp| U'_c}=2.4\times 10^4 \,{\rm m/s}.
\end{align}
Assuming a width $W=5\,{\rm \mu m}$ and $T=1\,$K the thermal conductance of the Goldstone mode in the CAF phase can be estimated from Eq.~(\ref{stefan-boltzmann}) as $G_{\rm th}\simeq 17\,$pW/K.

\section{Calculation of the particle-hole excitation spectrum}

We shall perform the calculation in several steps. We first evaluate the matrix elements of the Hamiltonian in the single-particle eigenbasis of the various broken-symmetry phases of bilayer graphene. This is followed by an evaluation of the single-particle self energy in Sec.~\ref{sec:self_energy}. Finally, we evaluate the response function in Sec.~\ref{sec:response_function} and find the excitation spectrum from its poles.

\subsection{Matrix elements in SU(4) spin space}

As introduced in the main text we use a notation, where the eight states in the zero-energy Landau level are labeled by index $a=\{n_a,\alpha\}$ that combines orbital index $n_a=0,1$ and SU(4) index $\alpha$. The Landau level wavefunctions are $\psi_{a,k}({\bf r})=\chi_{\alpha}\braket{{\bf r}|n_a,k}$, where $\chi_{\alpha}$ is an SU(4) spinor and $\ket{n,k}$ is the orbital wavefunction.

\subsubsection{Eigenbases of broken-symmetry states}\label{sec:eigenbases}
The single particle Hamiltonian ${\cal H}_s=-\epsilon_Z\sigma_z-\epsilon_V\tau_z$ is given in the spinor basis $\chi^0=(\chi_{K,\uparrow},\chi_{K,\downarrow},\chi_{K',\uparrow},\chi_{K',\downarrow})^T$. The SU(4)-spin symmetric part of the Hamiltonian is minimized when the four electrons occupy the single-particle states $\chi_{\alpha}\ket{0}$, $\chi_{\alpha}\ket{1}$, $\chi_{\beta}\ket{0}$, $\chi_{\beta}\ket{1}$ \cite{Barlas2008,Abanin2009,Kharitonov2012} where $\alpha$ and $\beta$ are SU(4) indices. Taking into account the short-range anisotropic interactions, the system can be in four different phases: CAF, FM, PLP, or FLP. These four phases can be described by two classes of single-particle eigenstates \cite{Kharitonov2012,Kharitonov2012a}.

(a) {\em CAF and FM.} In the canted antiferromagnet phase, we have $\alpha=\{K,{\bf s}_\alpha\}$ and $\beta=\{K',{\bf s}_\beta\}$ with the spin polarization vectors ${\bf s}_{\alpha/\beta}=(\pm\sin\theta_s\cos\varphi_s,\pm \sin\theta_s\sin\varphi_s,\cos\theta_s)$. The spinor $\chi^{\rm CAF}=(\chi_{K,{\bf s}_\alpha},\chi_{K,-{\bf s}_{\alpha}},\chi_{K',{\bf s}_\beta},\chi_{K',-{\bf s}_{\beta}})^T$ is related to $\chi^0$ by the transformation $ \chi^{\rm CAF}=U_{\rm CAF}^\dag \chi^0$ with
\begin{align}
U_{\rm CAF}=e^{i\sigma_y\tau_z\theta_s/2}e^{i\sigma_z\varphi_s/2}.
\end{align}
The ferromagnetic phase is realized for $\theta_s=0$.

(b) {\em PLP and FLP.} In the partially layer polarized state we have $\alpha=\{{\bf m},\uparrow\}$ and $\beta=\{{\bf m},\downarrow\}$ with the isospin polarization vector ${\bf m}=(\sin\theta_m\cos\varphi_m, \sin\theta_m\sin\varphi_m,\cos\theta_m)$. 
The spinor $\chi^{\rm PLP}=(\chi_{{\bf m},\uparrow},\chi_{{\bf m},\downarrow},\chi_{-{\bf m},\uparrow},\chi_{-{\bf m},\downarrow})^T$ is related to $\chi^0$ by the transformation $ \chi^{\rm PLP}=U_{\rm PLP}^\dag \chi^0$ with
\begin{align}
U_{\rm PLP}=e^{i\tau_y\theta_m/2}e^{i\tau_z\varphi_m/2}.
\end{align}
The fully layer polarized phase is realized for $\theta_m=0$.

\subsubsection{Useful expressions}
In the CAF and PLP eigenbases the single-particle Hamiltonian ${\cal H}_s=-\epsilon_V\tau_z-\epsilon_Z\sigma_z$ is given by
\begin{align}
{\cal H}_s^{\rm CAF}=& U^\dag_{\rm CAF} {\cal H}_s U_{\rm CAF}=-\epsilon_V\tau_z-\epsilon_Z (\sigma_z \cos\theta_s +\tau_z\sigma_x\sin\theta_s\cos\varphi_s+\tau_z\sigma_y\sin\theta_s\sin\varphi_s)\label{Hs_caf},\\
{\cal H}_s^{\rm PLP}=&-\epsilon_V (\tau_z \cos\theta_m +\tau_x\sin\theta_m\cos\varphi_m+\tau_y\sin\theta_m\sin\varphi_m)-\epsilon_Z\sigma_z. \label{Hs_plp}
\end{align}

It is useful to evaluate certain expressions involving the anisotropic interaction $W_{\alpha\beta\gamma\delta}$ in the eigenbases that will be needed later. For the evaluation of the self energy in Sec.~\ref{sec:self_energy} we will need the $4\times 4$ matrix 
\begin{align}
 N_{\alpha\beta}=\frac{1}{\pi l_B^2}\sum_{\lambda}f_\lambda ( W_{\alpha\lambda\beta\lambda}-W_{\alpha\lambda\lambda\beta}) ,
\end{align}
where $f_\lambda$ is the occupation number. In the eigenbases we find
\begin{align}
N^{\rm CAF}_{\alpha\beta} =&   u_z\delta_{\alpha\beta}f_\alpha+2u_\perp\left[\frac{1}{2}(1-\sigma_z)
+\cos\theta_s(\sigma_z\cos\theta_s+\tau_z\sigma_x\sin\theta_s\cos\varphi_s+\tau_z\sigma_y\sin\theta_s\sin\varphi_s) \right]_{\alpha\beta}\label{W_selfenergy_CAF},\\
N^{\rm PLP}_{\alpha\beta} =&  \delta_{\alpha\beta}(1-f_\alpha)(2u_\perp+u_z)-(u_z\cos^2\theta_m+u_\perp\sin^2\theta_m)(\tau_z)_{\alpha\beta}\nonumber\\
&\qquad +(u_\perp-u_z)\cos\theta_m\sin\theta_m(\cos\varphi_m\tau_x+\sin\varphi_m\tau_y)_{\alpha\beta}\label{W_selfenergy_PLP}.
\end{align}

During the evaluation of the response function in Sec.~\ref{sec:response_function}, $W$ appears in the combination 
\begin{align}
M_{\alpha\beta;\lambda\mu}= \frac{W_{\alpha\mu\lambda\beta}-W_{\alpha\mu\beta\lambda}}{\pi l_B^2}\label{define_Wmatrix}
\end{align}
for all pairs $(\alpha,\beta)$ and $(\lambda,\mu)$ such that each pair includes one occupied and one empty state. For each pair there are eight possible combinations and thus we can write $M$ as an $8\times 8$ matrix. For simplicity we set $\phi_m=\phi_s=0$ as the spectrum does not depend on the azimuthal angle.

In the CAF and FM phase we chose the basis  $\ket{\lambda\mu}=\chi^{\rm CAF}_\lambda \otimes\chi^{\rm CAF}_\mu$. Thus $\ket{\lambda}$ is occupied for $\lambda=1,3$ and empty for $\lambda=2,4$. Writing the matrix $M$ in the basis $(\ket{12},\ket{21},\ket{34},\ket{43},\ket{14},\ket{23},\ket{32},\ket{41})$ we find after some calculation
\begin{align}
 M_{\rm CAF}=-u_z +\begin{pmatrix}
    0&A_1&0&0\\
    A_1&0&0&0\\
    0&0&A_2&0\\
    0&0&0&A_2
   \end{pmatrix}\label{CAF_Wmatrix},
\end{align}
where we have defined the $2\times 2$ matrices $A_1=2u_\perp(-\cos^2\theta_s+\rho_x\sin^2\theta_s)$ and $A_2=2u_z+u_\perp-u_\perp\cos2\theta_s-2\rho_x u_\perp \sin^2\theta_s$ and $\rho_x$ is a Pauli matrix.

In the PLP and FLP phase we chose the basis $\ket{\lambda\mu}=\chi^{\rm PLP}_\lambda \otimes\chi^{\rm PLP}_\mu$. In this case $\ket{\lambda}$ is occupied for $\lambda=1,2$ and empty for $\lambda=3,4$. We obtain $M$ in the basis $(\ket{13},\ket{31},\ket{24},\ket{42},\ket{14},\ket{32},\ket{23},\ket{41})$ as
\begin{align}
 M_{\rm PLP}=2u_\perp+u_z +\begin{pmatrix}
    0&B_1&0&0\\
    B_1&0&0&0\\
    0&0&B_2&0\\
    0&0&0&B_2
   \end{pmatrix}\label{PLP_Wmatrix}
\end{align}
with $2\times 2$ matrices $B_1= (3 u_\perp + u_z + (u_\perp - u_z) \cos2 \theta_m)/2+( u_z-u_\perp) \sin^2\theta_m\rho_x$ and  $B_2= u_z\cos^2\theta_m+u_\perp\sin^2\theta_m-2u_\perp-u_z -( u_z-u_\perp) \sin^2\theta_m\rho_x$.

\subsection{Interaction matrix elements in orbital space}

The Coulomb interaction can be decomposed into a long-range isotropic and a short-range anisotropic contribution
\begin{align}
 V_{abcd}(q_1,q_2,q_3,q_4)=\int d^2r_1d^2r_2\psi_{a,q_1}^*({\bf r}_1) \psi^*_{b,q_2}({\bf r}_2) \psi_{c,q_3}({\bf r}_2)\psi_{d,q_4}({\bf r}_1) V_{\alpha\beta\gamma\delta}({\bf r}_1-{\bf r}_2)\label{interaction},
\end{align}
where $V_{\alpha\beta\gamma\delta}({\bf r})=V_0({\bf r})\delta_{\alpha \delta}\delta_{\beta \gamma}+\delta({\bf r})W_{\alpha \beta \gamma \delta}$. The long-range Coulomb interaction $V_0({\bf r})=e^2/\epsilon r$ includes an effective dielectric constant, which takes into account screening by electrons, the substrate and nearby gates. The short-range anisotropic interaction is
\begin{align}
W_{\alpha \beta \gamma \delta}=\pi l_B^2\sum_{j=x,y,z} u_j(\tau_j)_{\alpha\delta}(\tau_j)_{\beta\gamma}\label{W_matrix}.
\end{align}
After introducing relative and center of mass coordinates, $x_{1/2}=X\pm x/2$ and $y_{1/2}=Y\pm y/2$, and shifting $X\to X+(q_1+q_2)l_B^2/2$ we find
\begin{align}
 V_{abcd}(q_1,q_2,q_3,q_4)=&\delta_{q_1+q_2,q_3+q_4}\tilde{V}_{abcd}(q_1-q_3,q_2-q_3)\label{Vtilde}
\end{align}
with
\begin{align}
\tilde{V}_{abcd}(q,q')= &\frac{ e^{-q'^2l_B^2/2}}{\pi l_B^2L_y\left(2^{n_a+n_b+n_c+n_d}\right)^{1/2}}\int dxdy\, e^{-(x+ql_B^2)^2/2l_B^2}  e^{-iyq'}V_{\alpha\beta\gamma\delta}({\bf r})
 \int dX e^{-2X^2/l_B^2} \nonumber\\
 &\times H_{n_a}(X+\frac{x}{2}+(q-q')l_B^2/2)H_{n_b}(X-\frac{x}{2}-(q-q')l_B^2/2)
 H_{n_c}(X-\frac{x}{2}-(q+q')l_B^2/2)\nonumber\\
 &\times H_{n_d}(X+\frac{x}{2}+(q+q')l_B^2/2) .
\end{align}
where we have used the Landau level wavefunction
\begin{align}
\braket{{\bf r}|n,k}=e^{iky}\frac{1}{\sqrt{L_yl_B 2^n\sqrt{\pi}}}H_n[(x+kl_B^2)/l_B]e^{-(x+kl_B^2)^2/2 l_B^2}
\end{align}
with $H_0(x)=1$ and $H_1(x)=x$. We will need the partial Fourier transforms
\begin{align}
 \tilde{V}^{(1)}_{abcd}({\bf k})&=\sum_{q'}e^{iq'k_xl_B^2}\tilde{V}_{abcd}(k_y,q')\label{V_1},\\
 \tilde{V}^{(2)}_{abcd}({\bf k})&=\sum_{q}e^{iqk_xl_B^2}\tilde{V}_{abcd}(q,k_y)\label{V_2}.
\end{align}
The isotropic parts read \cite{Kallin1984}
\begin{align}
 \tilde{V}^{(1,i)}_{abcd}({\bf k})
 =&\delta_{\alpha \delta}\delta_{\beta \gamma}
 \sqrt{\frac{2^{n_b+n_c}}{2^{n_a+n_d}}}
\int  \frac{d^2r}{2\pi l_B^2} e^{-r^2/2l_B^2} V_0({\bf r}-{\bf k}\times \hat{z}l_B^2) \left(\frac{x+iy}{l_B}\right)^{n_a-n_c}
\left(\frac{x-iy}{l_B}\right)^{n_d-n_b}\nonumber\\
&\times L_{n_c}^{n_a-n_c}(r^2/2l_B^2)
L_{n_b}^{n_d-n_b}(r^2/2l_B^2)
\end{align}
and
\begin{align}
 \tilde{V}^{(2,i)}_{abcd}({\bf k})
 =&\delta_{\alpha \delta}\delta_{\beta \gamma}
 \sqrt{\frac{2^{n_b+n_d}}{2^{n_a+n_c}}}
\int\frac{ d^2r }{2\pi l_B^2}e^{-i{\bf kr}}V_0({\bf r})l_B^{n_a-n_d+n_c-n_b} \left(ik_x-k_y\right)^{n_a-n_d}
\left(-ik_x-k_y\right)^{n_c-n_b}\nonumber\\
&\times L_{n_d}^{n_a-n_d}(k^2l_B^2/2)
L_{n_b}^{n_c-n_b}(k^2l_B^2/2)e^{-k^2l_B^2/2}.
\end{align}
The anisotropic parts are
\begin{align}
 \tilde{V}^{(1,a)}_{abcd}({\bf k})
 =& \sqrt{\frac{2^{n_b+n_c}}{2^{n_a+n_d}}}
  \frac{W_{\alpha\beta\gamma\delta}}{2\pi l_B^2}  l_B^{n_a-n_c+n_d-n_b}\left(k_y-ik_x\right)^{n_a-n_c}
\left(k_y+ik_x\right)^{n_d-n_b}\nonumber\\
&\times L_{n_c}^{n_a-n_c}(k^2l_B^2/2)
L_{n_b}^{n_d-n_b}(k^2l_B^2/2) e^{-k^2l_B^2/2}
\end{align}
and
\begin{align}
 \tilde{V}^{(2,a)}_{abcd}({\bf k})
 =& \sqrt{\frac{2^{n_b+n_d}}{2^{n_a+n_c}}}
\frac{W_{\alpha\beta\gamma\delta}}{2\pi l_B^2}l_B^{n_a-n_d+n_c-n_b} \left(ik_x-k_y\right)^{n_a-n_d}
\left(-ik_x-k_y\right)^{n_c-n_b}\nonumber\\
&\times L_{n_d}^{n_a-n_d}(k^2l_B^2/2)
L_{n_b}^{n_c-n_b}(k^2l_B^2/2)e^{-k^2l_B^2/2}.
\end{align}
Here the $L_{n}^m$ are the associated Laguerre polynomials, $L^0_0(x)=L^1_0(x)=1$ and $L^0_1(x)=1-x$. Note that the order $m$ of $L_n^m$ should always be non-negative in the above expressions and indices should be interchanged if necessary. 
We can rewrite these results by defining the form factors 
\begin{align}
 F_{00}({\bf k})=&1,  &F_{11}({\bf k})=&1-\frac{k^2l_B^2}{2},\\
 F_{01}({\bf k})=&\frac{1}{\sqrt{2}}(k_y+ik_x)l_B, &F_{10}({\bf k})=&[F_{01}({\bf k})]^*.
\end{align}
The Hartree contribution of the isotropic interaction will not be needed as we only consider intra Landau level excitations. The Fock contribution reads
\begin{align}
 V_{ambl}^{(1,i)}({\bf r})=\delta_{\alpha\lambda}\delta_{\beta\mu}e^2\int \frac{d^2r}{2\pi l_B^2}\frac{\exp(-r^2/2l_B^2)}{|{\bf r}-l_B^2{\bf k}\times \hat{z}|}F_{ab}(\hat{z}\times{\bf r}/l_B^2)[F_{lm}(\hat{z}\times{\bf r}/l_B^2)]^*.\label{Viso}
\end{align}
The integrals are elementary and the solutions can expressed in terms of Bessel functions.
The anisotropic interaction will appear as a difference of Fock and Hartree terms
\begin{align}
  \tilde{V}^{(1,a)}_{ambl}({\bf k})-
 \tilde{V}^{(2,a)}_{amlb}(-{\bf k})=\frac{1}{2\pi l_B^2}(W_{\alpha\mu\lambda\beta}-W_{\alpha\mu\beta\lambda})e^{-k^2l_B^2/2}F_{ab}({\bf k})[F_{lm}({\bf k})]^*\label{Vani}.
\end{align}

\subsection{Self energy}\label{sec:self_energy}

The single-particle Green function is given by 
\begin{align}
 G_a(\omega)=\frac{1}{\omega-(\Sigma+{\cal H}_s)_a+i\epsilon_a}\label{sp_green_function}
\end{align}
with $\epsilon_a<0$ ($\epsilon_a>0$) if the state $a$ is occupied (empty). Here ${\cal H}_s$ is the single particle Hamiltonian and $\Sigma$ the self energy. We work in an eigenbasis such that the Green function is diagonal. Thus we require the sum $\Sigma+{\cal H}_s$ to be diagonal, although the self energy $\Sigma$ may have offdiagonal elements. 

In order to calculate the self energy we need to sum the Hartree and Fock contributions from anisotropic short-range interactions and isotropic long-range interactions. The long-range Coulomb interaction has only a Fock contribution, since the Hartree part is canceled by the ionic background. We find
\begin{align}
 \Sigma_{a b }(\omega,q)=&i\sum_{c}\int \frac{d\omega'}{2\pi}\sum_{q'}  G_{a}(\omega+\omega')e^{i(\omega+\omega')\delta}V_{a c b c }^i(q,q',q,q')\nonumber\\
 &+i\sum_{c }\int \frac{d\omega'}{2\pi}\sum_{q'}  G_{c }(\omega+\omega')e^{i(\omega+\omega')\delta}V^a_{a c b c }(q,q',q,q')\nonumber\\
 &-i\sum_{c }\int \frac{d\omega'}{2\pi}\sum_{q'}  G_{c }(\omega+\omega')e^{i(\omega+\omega')\delta}V^a_{a c c b }(q,q',q',q).
\end{align}
It is easy to see that the self energy does not depend on energy or momentum and we obtain
\begin{align}
 \Sigma_{a b }=&-\sum_{c }f_c 
 \left[\tilde{V}_{a c b c }^{(i,1)}(0)+\tilde{V}^{(a,1)}_{a c b c }(0)-\tilde{V}^{(a,2)}_{a c c b }(0)\right],
\end{align}
where $f_c $ is the occupation number. At this point we must proceed with care. The sum over intermediate states $c$ should run over all occupied Landau levels with $n_c\leq -2$ in addition to the partially filled zero-energy Landau level $n=0,1$. The contribution of the fully occupied Landau levels induces a ``Lamb shift'' between the $n=0$ and $n=1$ state \cite{Shizuya2012}. For the anisotropic contribution $\sum_{n_c\leq -2,\gamma}\tilde{V}^{(a,1)}_{a c b c }(0)-\tilde{V}^{(a,2)}_{a c c b }(0)$ yields just a constant energy shift that we can safely ignore. The isotropic part, however, depends on the orbital index $n_a$. Following Ref.~\cite{Shizuya2012} we can rewrite the sum over completely filled Landau levels as
\begin{align}
\sum_{n_c\leq2,\gamma} f_c\tilde{V}_{a c b c }^{(i,1)}(0)=-\frac{1}{2}\delta_{\alpha\beta}\sum_{n_c=0,1}\tilde{V}_{a c b c }^{(i,1)}(0).
\end{align}
Using Eqs.~(\ref{W_selfenergy_CAF}), (\ref{W_selfenergy_PLP}), (\ref{Vani}), and (\ref{Viso}) and introducing  $U_c=\sqrt{\pi/2}e^2/\epsilon l_B$ we find the self energy in the eigenbases introduced in Sec.~\ref{sec:eigenbases} 
\begin{align}
 \Sigma^{\rm CAF}_{ab}=&\delta_{a b}(1/2-f_\alpha)
U_c\left(\frac{3}{2}-n_a\frac{1}{4}\right)
 -\delta_{ab}f_\alpha u_z\nonumber\\
&-\delta_{n_a n_b}\left[
u_\perp (1-\sigma_z)
+2u_\perp \cos\theta_s(\sigma_z\cos\theta_s+\tau_z\sigma_x\sin\theta_s\cos\varphi_s+
\tau_z\sigma_y\sin\theta_s\sin\varphi_s) \right]_{\alpha\beta}
\end{align}
and
\begin{align}
 \Sigma^{\rm PLP}_{ab} =&\delta_{ab}(1/2-f_\alpha)
 U_c\left(\frac{3}{2}-n_a\frac{1}{4}\right)-\delta_{ab}(1-f_\alpha)(2u_\perp+u_z)\nonumber\\
&+\delta_{n_a n_b}[(u_z\cos^2\theta_m+u_\perp\sin^2\theta_m)\tau_z+(u_z-u_\perp)\cos\theta_m\sin\theta_m(\cos\varphi_m\tau_x+\sin\varphi_m\tau_y)]_{\alpha\beta}.
\end{align}
Note that the self energies are diagonal in orbital space. The single-particle Hamiltonian is given by Eqs.~(\ref{Hs_caf}) and (\ref{Hs_plp}).
In the CAF basis, we find that $\Sigma+{\cal H}_s$ is diagonal if $\cos\theta_s=-\epsilon_Z/2u_\perp$ (CAF) or $\theta_s=0$ (FM) whereas in the PLP basis we find $\cos\theta_m=\epsilon_V/(u_z-u_\perp)$ (PLP) or $\theta_m=0$ (FLP) in agreement with Ref.~\cite{Kharitonov2012a}
\begin{align}
(\Sigma+{\cal H}_s)^{\rm CAF} =&\delta_{ab}(1/2-f_\alpha) U_c\left(\frac{3}{2}-n_a\frac{1}{4}\right)-\delta_{ab}(f_\alpha u_z+u_\perp)\nonumber\\ &+\left[-\epsilon_V\tau_z+u_\perp\sigma_z-\sigma_z\cos\theta_s(2u_\perp\cos\theta_s+\epsilon_Z )\right]_{\alpha\beta}\label{self_energy_caf},\\
(\Sigma+{\cal H}_s)^{\rm PLP} =&\delta_{ab}(1/2-f_\alpha) U_c\left(\frac{3}{2}-n_a\frac{1}{4}\right)+\delta_{ab}[f_\alpha (2u_\perp+u_z)-(u_\perp+u_z)]\nonumber\\
&+\left[-\epsilon_Z\sigma_z- \tau_z \cos\theta_m[ \epsilon_V-(u_z-u_\perp)\cos\theta_m]\right]_{\alpha\beta}.
\label{self_energy_plp}
\end{align}

\subsection{Response function}\label{sec:response_function}

We now calculate the response function of operator $\theta_A$
\begin{align}
\chi_A({\bf k},\omega)= \sum_{ab}\sum_{q_1q_2}\Braket{q_1a|e^{i{\bf kr}}\theta_A^\dag|q_2b}\int \frac{d\omega'}{2\pi}G_a(\omega+\omega')G_b(\omega')\Gamma_{A,ab}(q_1,q_2;\mathbf{k},\omega),
\end{align}
closely following \cite{Kallin1984}. The single-particle Green function is given by Eq.~(\ref{sp_green_function}) and the vertex part can be determined from the diagrams shown in Fig.~\ref{fig:setup}(b) of the main text as
\begin{align}
 \Gamma_{A,ab}(q_1,q_2;{\bf k},\omega)=&\Braket{q_2b|e^{-i{\bf kr}}\theta_A|q_1a}\nonumber\\
 &+i
 \sum_{lm}\sum_{q_3q_4} V_{ambl}(q_1,q_4,q_2,q_3)\int \frac{d\omega'}{2\pi}G_l(\omega+\omega')G_m(\omega')\Gamma_{A,lm}(q_3,q_4,{\bf k},\omega)\nonumber\\
 &-i
 \sum_{lm}\sum_{q_3q_4} V_{amlb}(q_1,q_4,q_3,q_2)\int \frac{d\omega'}{2\pi}G_l(\omega+\omega')G_m(\omega')\Gamma_{A,lm}(q_3,q_4,{\bf k},\omega),
\end{align}
where $V_{abcd}(q_1,q_2,q_3,q_4)$ is given by Eq.~(\ref{interaction}).
We change variables to total and relative momentum $Q=(q_1+q_2)/2$ and $q=q_1-q_2$ and using Eq.~(\ref{Vtilde}) we write
\begin{align}
 \Gamma_{A,ab}(Q,q;{\bf k},\omega) =&  \delta_{q,k_y}
e^{(q_1+q_2+ik_x)^2l_B^2/4 -q_1^2l_B^2/2-q_2^2l_B^2/2}(\theta_A)_{ba}\nonumber\\
 &+i
 \sum_{lm}\sum_{q_4}\tilde{V}_{ambl}(q,q_4-Q)\int \frac{d\omega'}{2\pi}G_l(\omega+\omega')G_m(\omega')\Gamma_{A,lm}(q_4+q/2,q_4-q/2,{\bf k},\omega)\nonumber\\
 &-i
 \sum_{lm}\sum_{q_4} \tilde{V}_{amlb}(Q-q_4,-q) \int \frac{d\omega'}{2\pi}G_l(\omega+\omega')G_m(\omega')\Gamma_{A,lm}(q_4+q/2,q_4-q/2,{\bf k},\omega).
\end{align}
Introducing the partial Fourier transform
\begin{align}
 \tilde{\Gamma}_{A,ab}(p,q;{\bf k},\omega)=\sum_{Q} e^{ipQl_B^2}\Gamma_{A,ab}(Q,q;{\bf k},\omega)
\end{align}
we obtain
\begin{align}
 \tilde{\Gamma}_{A,ab}(p,q;{\bf k},\omega)  =&  \delta_{q,k_y}
\delta_{p,-k_x}e^{-(k_x^2+k_y^2)l_B^2/4}(\theta_A)_{ba}+i
 \sum_{lm}\tilde{V}^{(1)}_{ambl}(q,-p)\int \frac{d\omega'}{2\pi}G_l(\omega+\omega')G_m(\omega')\tilde{\Gamma}_{A,lm}(p,q,{\bf k},\omega)\\
 &-i
 \sum_{lm} \tilde{V}^{(2)}_{amlb}(p,-q) \int \frac{d\omega'}{2\pi}G_l(\omega+\omega')G_m(\omega')\tilde{\Gamma}_{A,lm}(p,q,{\bf k},\omega),
\end{align}
where $\tilde{V}^{(1/2)}$ have been defined in Eqs.~(\ref{V_1}) and (\ref{V_2}). When introducing
\begin{align}
 \Gamma'_{A,ab}({\bf k},\omega)=\delta_{q,k_y}
\delta_{p,-k_x}\tilde{\Gamma}_{A,ab}(p,q,{\bf k},\omega),
\end{align}
we can rewrite the matrix equation for the vertex part as
\begin{align}
 \Gamma'_{A,ab}({\bf k},\omega)  =& e^{-(k_x^2+k_y^2)l_B^2/4}(\theta_A)_{ba}+
 \sum_{lm}\left[\tilde{V}^{(1)}_{ambl}({\bf k})- \tilde{V}^{(2)}_{amlb}(-{\bf k})\right]
 D_{lm}(\omega) \Gamma'_{A,lm}({\bf k},\omega),
 \end{align}
where $D_{lm}$ is the two-particle propagator
\begin{align}
 D_{lm}(\omega)= &i \int \frac{d\omega'}{2\pi}G_l(\omega+\omega')G_m(\omega')\\
  =&\frac{f_l(1-f_m)}{\omega-[(\Sigma+{\cal H}_s)_l-(\Sigma+{\cal H}_s)_m]-i\epsilon}
-\frac{f_m(1-f_l)}{\omega-[(\Sigma+{\cal H}_s)_l-(\Sigma+{\cal H}_s)_m]+i\epsilon}.
\end{align}
Using Eqs.~(\ref{self_energy_caf}) and (\ref{self_energy_plp}) we obtain the exchange energy $ E_{{\rm xc},lm}=(\Sigma+{\cal H}_s)_{l}- (\Sigma+{\cal H}_s)_{m}$ for $l$ occupied and $m$ empty as
\begin{align}
E^{\rm CAF}_{{\rm xc},lm}=&- U_c\left[\frac{3}{2}-\frac{1}{8}(n_l+n_m)\right]
+2u_\perp-u_z-\epsilon_V[(\tau_z)_{ll}-(\tau_z)_{mm}]-2\cos\theta_s(2u_\perp\cos\theta_s+\epsilon_Z )
 \label{E_xc_caf},\\
E^{\rm PLP}_{{\rm xc},lm}=&- U_c\left[\frac{3}{2}-\frac{1}{8}(n_l+n_m)\right]+2u_\perp+u_z
-\epsilon_Z[(\sigma_z)_{ll}-(\sigma_z)_{mm}]
-2\cos\theta_m[ \epsilon_V-(u_z-u_\perp)\cos\theta_m]
 \label{E_xc_plp}
\end{align}
and thus
\begin{align}
 D_{lm}(\omega)
  =&\frac{f_l(1-f_m)}{\omega-E_{{\rm xc},lm}-i\epsilon}
-\frac{f_m(1-f_l)}{\omega+E_{{\rm xc},ml}+i\epsilon}.
\end{align}
The equation for the vertex part can be rewritten as
\begin{align}
 \sum_{lm}\left[\delta_{al}\delta_{bm} (D_{lm})^{-1} -\tilde{V}^{(1)}_{ambl}({\bf k})+ \tilde{V}^{(2)}_{amlb}(-{\bf k})\right]
\Pi_{Alm}({\bf k},\omega)
=& e^{-(k_x^2+k_y^2)l_B^2/4}(\theta_A)_{ba},
 \end{align}
where we have introduced
\begin{align}
 \Pi_{Alm}({\bf k},\omega)= D_{lm}(\omega) \Gamma'_{A,lm}({\bf k},\omega).
\end{align}
This equation must be satisfied for all pairs  $(a,b)$ such that one state is occupied and one is empty. The excitation spectrum is given by poles $\omega(k)$ of the response function which are solutions of the equation
\begin{align}
 \sum_{lm}\left[\delta_{al}\delta_{bm} (D_{lm})^{-1} -\tilde{V}^{(1)}_{ambl}({\bf k})+ \tilde{V}^{(2)}_{amlb}(-{\bf k})\right]
B_{lm}=0,
 \end{align}
where the summation runs over particle-hole pairs $(l,m)$ such that one state is occupied and one is empty, since $\Pi_{lm}$ vanishes otherwise. Defining $\tilde{V}_{ablm}=\tilde{V}^{(1)}_{ambl}+ \tilde{V}^{(2)}_{amlb}$ we arrive at Eq.~(\ref{eval_problem}) of the main text
\begin{align}
 \sum_{lm}\left[\delta_{al}\delta_{bm} 
 [(f_l-f_m)\omega-E_{{\rm xc},lm}] -\tilde{V}^{(1)}_{ambl}({\bf k})+ \tilde{V}^{(2)}_{amlb}(-{\bf k})\right]
B_{lm}=0\label{eval_problem_app}.
 \end{align}
Note that the Hartree part of the long-range interaction $\tilde{V}^{(2,i)}_{amlb}({\bf k})\propto \delta_{\alpha\beta}\delta_{\lambda\mu}$ vanishes, as we only consider intra-Landau level excitations for which $\alpha\neq\beta$ and $\lambda\neq \mu$.

\subsection{Low-energy excitation spectrum}

We now derive the spectrum at low-energies, $\omega\ll U_c$, and long-wavelengths, $kl_B\ll1$, in the case that long-range Coulomb interaction dominates all other scales $U_c\gg u_z,u_\perp,\epsilon_V,\epsilon_Z$. Equation~(\ref{eval_problem_app}) can be written as a matrix in a 32-dimensional space with four components in orbital space and eight components in SU(4) spin space of particle hole pairs $(a,b)$.

We start by diagonalizing the eigenvalue problem in four dimensional orbital space within the basis $\ket{n_an_b}=[\ket{00},\ket{01},\ket{10},\ket{11}]_{ab}$. The Hartree and Fock contributions of the anisotropic interaction are given by  Eq.~(\ref{Vani}) as
\begin{align}
  \tilde{V}^{(1,a)}_{ambl}({\bf k})-
 \tilde{V}^{(2,a)}_{amlb}(-{\bf k})=\frac{1}{2\pi l_B^2}(W_{\alpha\mu\lambda\beta}-W_{\alpha\mu\beta\lambda})e^{-k^2l_B^2/2}
F_{n_a,n_b}({\bf k})F_{n_l,n_m}({\bf k}).
\end{align}
In orbital space this expression has only one nonzero eigenvalue with the corresponding eigenvector  $\nu({\bf k})=[F_{00}({\bf k}),F_{01}({\bf k}),F_{10}({\bf k}),F_{11}({\bf k})]$. Thus the anisotropic contribution vanishes outside the subspace spanned by the vector $\nu({\bf k})$.

We now turn to the remaining part $\delta_{ac}\delta_{bd} E_{{\rm xc},cd} +\tilde{V}^{(1,i)}_{ambl}({\bf k})$. At $k=0$ and only keeping terms of order $U_c$, this matrix is an identity matrix in SU(4) spin space and has eigenvalues $U_c$, $7U_c/8$, $7U_c/8$, and 0 in orbital space. Incidentally the zero-eigenvalue subspace is spanned by the vector $\nu({\bf k}=0)=(1,0,0,1)$. Hence, the low-energy excitations at $k=0$ comprise superpositions of electrons and holes in the same orbital state $\ket{00}+\ket{11}$.

The vector $\nu({\bf k})$ therefore defines the low-energy subspace near $k=0$ with four modes and their particle-hole symmetric counterparts. We project onto this subspace using a Schrieffer-Wolff transformation and obtain an effective $8\times8$ eigenvalue problem in SU(4) spin space expanded to second order in $kl_B$
\begin{align}
 \sum_{lm}\left[\delta_{al}\delta_{bm} 
 [(f_l-f_m)\omega+U'_ck^2l_B^2
 -\Delta_{lm} -\frac{1-k^2l_B^2/2}{2\pi l_B^2}(W_{\alpha\mu\lambda\beta}-W_{\alpha\mu\beta\lambda})\right]
B_{lm}=0\label{eval_final}
 \end{align}
with $U'_c=N_0U_c$, $N_0=89/224$, and 
\begin{align}
\Delta^{\rm CAF}_{lm}=&2u_\perp-u_z-(f_l-f_m)\epsilon_V[(\tau_z)_{ll}-(\tau_z)_{mm}]-2\cos\theta_s(2u_\perp\cos\theta_s+\epsilon_Z ),\\
\Delta^{\rm PLP}_{lm}=&2u_\perp+u_z
-(f_l-f_m)\epsilon_Z[(\sigma_z)_{ll}-(\sigma_z)_{mm}]
-2\cos\theta_m[ \epsilon_V-(u_z-u_\perp)\cos\theta_m] .
\end{align}
The anisotropic contribution $W_{\alpha\mu\lambda\beta}-W_{\alpha\mu\beta\lambda}$ is given by Eqs.~(\ref{define_Wmatrix})--(\ref{PLP_Wmatrix}). Equation~(\ref{eval_final}) contains the solutions in all four phases depending on the choice of basis and the definitions of $\theta_{s/m}$. 
The $8\times8$ matrix written in the basis used in Eq.~(\ref{CAF_Wmatrix}) and (\ref{PLP_Wmatrix}) decomposes into a $4\times 4$ block and two $2\times 2$ blocks, which can be readily diagonalized. The solutions in the CAF and PLP phase are given as Eqs.~(\ref{spin_goldstone})--(\ref{PLP2}) of the main text.

In the FM phase we can set $\theta_s=0$ and find
\begin{align}
 \omega_1=&2\epsilon_Z+k^2l_B^2 U_c',\\
 \omega_2=&2(\epsilon_Z+2u_\perp)+k^2l_B^2 U_c',\\
 \omega_{3,4}=&2(\epsilon_Z+u_z+u_\perp\pm \epsilon_V)+k^2l_B^2 U_c'.
\end{align}
We see that the gap of all four low-energy modes increases as the Zeeman field increases. Gap closings occur for  $\epsilon_V=-2u\perp$ at the CAF-FM transitions and for $\epsilon_V-\epsilon_Z=u_z+u_\perp$ at the FM-FLP transition \cite{Kharitonov2012a}.

In the FLP phase we set $\theta_p=0$ and obtain
\begin{align}
 \omega_{1/2}=&2(\epsilon_V-u_z\pm u_\perp)+k^2l_B^2 U_c',\\
 \omega_{3/4}=&2(\epsilon_V-u_z-u_\perp\pm\epsilon_Z)+k^2l_B^2 U_c'.
\end{align}
As in the FM phase all modes have a gap, which increases with $\epsilon_V$. A gap closing occurs at the PLP-FLP transition, where $\epsilon_V=u_z-u_\perp$, and the FM-FLP transition, where $\epsilon_V-\epsilon_Z=u_z+u_\perp$ \cite{Kharitonov2012a}.

\end{document}